\documentclass[reprint,twocolumn,nofootinbib]{revtex4-1}
\usepackage{graphicx}%
\usepackage{dcolumn}%
\usepackage{bm}%
\usepackage{epsfig,colordvi}
\usepackage[latin1]{inputenc}
\usepackage{color} 
\usepackage{here}

 \definecolor{dgreen}{rgb}{0.03, 0.47, 0.19}

 \def\be{\begin{equation}}
 \def\ee{\end{equation}}
 \def\bea{\begin{eqnarray}}
 \def\eea{\end{eqnarray}}
 \def\bean{\begin{eqnarray*}}
 \def\eean{\end{eqnarray*}}
 \def\gsim{\mathrel{\rlap{\lower0.2em\hbox{$\sim$}}\raise0.2em\hbox{$>$}}}
 \def\ksim{\mathrel{\rlap{\lower0.2em\hbox{$\sim$}}\raise0.2em\hbox{$<$}}}
 \def\kg{\mathrel{\rlap{\lower0.25em\hbox{$>$}}\raise0.25em\hbox{$<$}}}

\makeatletter
\newsavebox\myboxA
\newsavebox\myboxB
\newlength\mylenA

\newcommand*\xoverline[2][0.75]{%
    \sbox{\myboxA}{$\m@th#2$}%
    \setbox\myboxB\null
    \ht\myboxB=\ht\myboxA%
    \dp\myboxB=\dp\myboxA%
    \wd\myboxB=#1\wd\myboxA
    \sbox\myboxB{$\m@th\overline{\copy\myboxB}$}
    \setlength\mylenA{\the\wd\myboxA}
    \addtolength\mylenA{-\the\wd\myboxB}%
    \ifdim\wd\myboxB<\wd\myboxA%
       \rlap{\hskip 0.5\mylenA\usebox\myboxB}{\usebox\myboxA}%
    \else
        \hskip -0.5\mylenA\rlap{\usebox\myboxA}{\hskip 0.5\mylenA\usebox\myboxB}%
    \fi}
\makeatother

\begin{document}

\title{Influence of final-state radiation on heavy-flavour observables in pp collisions}

\author{L. Vermunt$^1$, J. Aichelin$^{2}$, P.B. Gossiaux$^2$, K. Werner$^2$, A. Mischke$^1$, B. Guiot$^{2,3}$, M. Nahrgang$^2$ and T. Pierog$^4$ }
\affiliation{$^1$ Institute for Subatomic Physics, Utrecht University, Utrecht, the Netherlands}
\affiliation{$^2$ SUBATECH, \'Ecole des Mines de Nantes, Universit\'e de Nantes, IN2P3/CNRS, Nantes, France}
\affiliation{$^3$ University T\'ecnica Frederico Santa Maria, Valparaiso, Chile}
\affiliation{$^4$ Karlsruhe Institute of Technology, IKP, Karlsruhe, Germany}

\date{\today}

\begin{abstract} \noindent
The influence of final-state radiation (FSR) of heavy quarks on observables in high-energy proton-proton collisions is studied. The transverse momentum correlation of $\rm D$ and $\xoverline{\rm D}$ mesons, which have been emitted with an azimuthal difference angle close to 180 degrees, is identified as an observable which is sensitive to the FSR process. We demonstrate this by performing calculations with the EPOS3+HQ model and with the event generator Pythia 6. The initial symmetric $p_{\rm T} = p_{\rm T}{'}$ correlation for back-to-back pairs does not completely vanish in EPOS3+HQ, neither for the final $\rm D\xoverline{\rm D}$ pairs nor for the $c\bar{c}$ pairs before hadronisation. Also a significant difference in the shape of the correlation distribution for EPOS3+HQ and Pythia 6 is observed. Therefore, particle correlations in pp data offer the possibility to study several aspects of energy loss in heavy-ion collisions.
\end{abstract}


\maketitle

\section{Introduction}
For all high-energy processes that contain particles with colour and/or charge, initial- and final-state radiation are important processes to be taken into account. These hard radiation corrections become more important with increasing beam energy. So for the physics interpretation of the data at the Large Hadron Collider (LHC), it is important to have a good understanding of these processes. For simple cases like $e^{-}e^{+}\rightarrow \gamma^{*} / Z^{0} \rightarrow q\bar{q}g$, the matrix elements can still be calculated analytically. However, for most of the processes next-to-leading order (NLO) or even higher order processes become important and analytic calculations become impossible.

These perturbative QCD corrections can be modelled using Monte Carlo algorithms of the DGLAP equations \cite{Gribov:1972ri,Altarelli:1977zs,Dokshitzer:1977sg}. The DGLAP description introduces a probabilistic approximation with a branching probability that is independent of other sub-processes. This approach is the state-of-the-art, and most event generators have implemented it in a similar way. However, on the quantitative level, differences between the models arise due to different implementations of the DGLAP equations. For example, there are different possibilities in the way the incident partons are generated before they undergo a parton-shower (e.g. the choice of the initial maximal virtuality), in the interpretation of the kinematics of a branching and for a possible interaction with a medium. Because presently these questions are not solved, it is important to find observables, which are sensitive to the FSR process, which can then be compared to different model calculations.

To measure FSR of quarks is, however, not that easy. It is a process that takes place immediately after the production of coloured objects which, because of confinement, can not be measured independently from their environment. In this paper, a transverse momentum correlation observable is identified, which is sensitive to FSR. We will focus on heavy-flavour particles because these are most likely produced in pairs in the initial stages of the collision due to flavour conservation. Leading heavy mesons can also be easily identified. Up to now, traditional heavy-flavour observables, like the nuclear modification factor $R_{AA}$ and the elliptic flow parameter $v_{2}$, have been used extensively in heavy-ion physics. Measurements of these observables point to a significant in-medium energy loss of heavy quarks and a partial thermalisation of charm quarks with the medium \cite{Adam:2015sza,Sirunyan:2017oug,Abelev:2013lca}. With the upcoming detector upgrades for the LHC Run-3 \cite{Abelevetal:2014dna,CMS:2012sda}, statistically significant correlation measurements of heavy-flavour particles will also become feasible. Such pair observables contain very interesting information not only for the analysis of heavy-ion reactions \cite{Aarts:2016hap,Nahrgang:2013saa,Song:2016rzw,Mischke:2008af}, but also for proton-proton studies. Several correlation measurements of heavy-flavour particles were performed \cite{Acosta:2004nj,Khachatryan:2011wq,CMS:2016krr,Aaij:2012dz,Aaij:2017pvu}.

Our transverse momentum correlation observable was originally considered as an observable to differentiate in heavy-ion collisions between collisional and collisional plus radiative in-medium energy loss mechanisms. It turned out that this observable is as well sensitive to FSR. In pp collisions, where it is currently under debate whether a quark-gluon plasma (QGP) is produced with which the heavy quarks interact \cite{ALICE:2017jyt,Khachatryan:2016txc,Khachatryan:2010gv}, it will be almost exclusively sensitive to the final-state radiation process. For this study, we use the EPOS3+HQ model \cite{Werner:2013tya} and event generator Pythia 6 \cite{Sjostrand:2006za}. In both of these models we study the transverse momentum correlation of heavy quark pairs. In EPOS3+HQ, also the influence of a possibly produced medium can be investigated. In section~\ref{sec:FSR}, final-state radiation and the way this is usually modelled is discussed. In section~\ref{sec:Strategy}, the proposed strategy is presented. The results and conclusions will follow in sections~\ref{sec:Results} and \ref{sec:Conclusions}.

\section{Final-state radiation}
\label{sec:FSR}
The evolution of the final-state radiation shower, or more precise, the probability for a parton to branch, is given by the DGLAP evolution equations \cite{Gribov:1972ri,Altarelli:1977zs,Dokshitzer:1977sg}. These equations are based on the assumption that the cross section of a process that emits $n+1$ partons, $\sigma_{n+1}$, can be factorised into the cross section $\sigma_{n}$ and a probability density for the branching of one of these $n$ partons. The incident partons are characterised by their virtuality scale $Q^2$, where it is convenient to introduce
\begin{equation}
    t = \ln{(Q^2/\Lambda_{\rm QCD}^2)} \quad \Rightarrow \quad {\rm{d}} t = {\rm{d}} \ln{(Q^2)} = \frac{{\rm{d}} Q^2 }{ Q^2 }.
\end{equation}
The differential probability for a parton to branch, according to the DGLAP evolution, can then be written as:
\begin{equation}
    {\rm{d}}\mathcal{P}_{a} = \sum_{b,c} \frac{\alpha_{abc}}{2\pi} P_{a \rightarrow bc}(z) {\rm{d}}t {\rm{d}}z \, ,
    \label{eq:DGLAP}
\end{equation}
where the sum runs over all allowed branchings, $\alpha_{abc}$ is the corresponding coupling constant and $z$ is the energy fraction of the daughter partons. $P_{a \rightarrow bc}(z)$ are the so-called splitting functions, which can be calculated using standard collinear splitting for each type of branching:
\begin{eqnarray}
    P_{q\rightarrow qg}(z) &=& \frac{4}{3} \frac{1 + z^2}{1 - z} \, , \nonumber \\ 
    P_{g\rightarrow gg}(z) &=& 3 \frac{(1-z(1-z))^2}{z(1 - z)} \, , \\ 
    P_{g\rightarrow q\bar{q}}(z) &=& \frac{1}{2}(z^2 + (1-z)^2) \, . \nonumber
\end{eqnarray}

A Monte Carlo (MC) version of this DGLAP evolution has been implemented in several event generators, such as EPOS3 \cite{Werner:2013tya} and Pythia 6 \cite{Sjostrand:2006za}, which are studied in this paper. The properties of both models will be discussed in the next subsection. In Monte Carlo simulations of the DGLAP evolution, the simulated radiation showers start with a parton with a given initial virtuality $Q$ between zero and a fixed maximum $Q_{\rm max}$. The values for $Q^2$ are increasing for initial- and decreasing for final-state showers when the shower evolves. Equation \ref{eq:DGLAP} gives therefore the probability that during a change ${\rm{d}}Q^2$ of the virtuality, parton $a$ splits into daughters $b$ and $c$. To obtain the probability that an incoming parton splits at precisely this virtuality, the branching probability needs to be multiplied with the probability that the parton does not split at a higher virtuality. This is given by the Sudakov form factor $S(Q_{\rm max},Q)$. A cut-off parameter is introduced to terminate the parton shower when the virtuality becomes too high or too low. For FSR, this can be motivated as a parametrisation of the transition from perturbative splitting processes to non-perturbative hadronisation processes.

It is not clear how to define the maximum virtuality of the initial partons in FSR showers. The choice made for this upper scale $Q_{\rm max}$ can strongly affect the amount of well-separated jets \cite{phdmartin,Rohrmoser:2016yct}. Because these showers are time-like (the partons have $m^2 = E^2 - \mathbf{p}^2 \ge 0$) the virtuality $Q^2$ can be related to $m^2$ of the branching parton. Another possibility is to choose a $p_{\perp}$-ordered shower. Models differ in their choice for $Q_{\rm max}$ as well as in different cut-offs. In addition to this, the angular ordering of the parton cascade, necessary to treat the FSR process in a probabilistic way, is also differently realised. Usually, these various implementations are determined by performing some calibration on key observables in pp collisions, like the thrust or the humped back plateau for light hadrons.

\subsection{Models}

In this paper, a transverse momentum correlation observable is studied using EPOS3+HQ and Pythia 6. We will now shortly present both models and discuss the calibration in section \ref{subsect:Calibration}.

Our model EPOS3+HQ couples the Monte-Carlo treatment of the Boltzmann equation of heavy quarks (MC@sHQ) \cite{Gossiaux:2008jv} to the $3+1$ dimensional fluid dynamical evolution of the locally thermalised QGP following the initial conditions from EPOS3 \cite{Werner:2013tya}. EPOS3 is an universal event generator for pp, pA and AA collisions. These collisions are modelled on an event-by-event basis. The EPOS3 simulation, including the possible production of a medium, is based on the following stages:
\begin{itemize}
    \item \textbf{Initial conditions:} The Parton-Based Gribov-Regge Theory (PBGRT) \cite{Drescher:2000ha} is used for the multiple scattering approach. The elementary objects, so-called pomerons, are based on a DGLAP parton ladder. These parton ladders are treated in EPOS3 as classical relativistic strings.
    
    \item \textbf{Core-corona approach:} At some early proper time $\tau_0$, one separates fluid (core) and escaping hadrons, including jet hadrons (corona), based on the momenta and the density of string segments \cite{Werner:2007bf,Werner:2013tya}. The corresponding energy-momentum tensor of the core part is transformed into an equilibrium one, needed to start the hydrodynamical evolution. This is based on the hypothesis that equilibration happens rapidly and affects essentially the space components of the energy-momentum tensor.
    
    \item \textbf{Viscous hydrodynamic expansion:} Relativistic viscous hydrodynamic equations are used to evolve the core part of the system, starting from the initial proper time $\tau_{0}$ \cite{Werner:2013tya,Karpenko:2013wva}. For the shear velocity, $\eta /s = 0.08$ is taken. A cross-over equation-of-state is used, compatible with lattice QCD \cite{Borsanyi:2010cj,Werner:2010ny}.

    \item \textbf{Statistical hadronisation:} The thermal phase of the core ends when the hypersurface reaches the hadronisation temperature $T_{H}$ (taken in the region where the energy density varies strongly with temperature \cite{Werner:2010ny}). At this point, statistical hadronisation is employed to transform the ``core-matter'' into hadrons.
 
    \item \textbf{Final-state hadronic cascade:} After the statistical hadronisation, the hadron density is still high enough for hadronic scattering. The UrQMD model \cite{Bleicher:1999xi, Petersen:2008dd} is used to follow these hadronic scatterings until they cease.
\end{itemize}

Since the upgrade to EPOS3, heavy quarks ($Q$) are, just like the light quarks, produced in the initial stage using the PBGRT formalism. The parton ladders are composed of two space-like parton cascades and a Born process, which both can emit time-like partons (leading to time-like cascades). In all these processes, $Q\xoverline{Q}$ production is possible, which means that LO+NLO production mechanisms are taken into account in EPOS3+HQ. Also the modified kinematics in case of non-zero quark masses ($m_{c}=1.3$ and $m_{b}=4.2$~GeV/$c$ are used) is properly treated. A good agreement of $\rm D$-meson production in EPOS3 with available LHC data is found \cite{benjamin}. 

For the event generator Pythia, version 6.428 with the IBK-CTEQ5L Innsbruck tune is used. Because the majority of MC simulations for LHC experiments, especially for the very few heavy-flavour correlation measurements available, are using Pythia 6 tunes with a fair success with respect to experimental data, we also choose to use this relative old version. Pythia 8 does have some new features in its FSR machinery, like $\gamma \rightarrow q\bar{q}$ and $\gamma \rightarrow l^{+}l^{-}$ branchings and extensions to handle bremsstrahlung in Hidden Valley models \cite{Sjostrand:2014zea}, but none of them are essential for this study. Recent LHC tunes in Pythia are using $p_{\perp}$-ordered showers, where EPOS3 uses virtuality ordering.

Pythia 6 is using the Donnachie and Landshoff parametrisation to calculate the total hadronic cross section for $AB \rightarrow \rm{anything}$, which appears as the sum of a pomeron and a reggeon term \cite{Donnachie:1992ny}. The total cross section is subdivided into elastic, single diffractive, double diffractive and non-diffractive components. The first three components are described according to the model of Schuler and Sj\"ostrand based on Regge theory \cite{Schuler:1993td,Schuler:1993wr}, where the latter is given by ``whatever is left'' \cite{Sjostrand:2006za}. Once the process is selected, the kinematic variables are determined using the Lund string model \cite{Andersson:1983ia}. The default option in Pythia for heavy-flavour production is without taking the heavy-quark masses into account. The mass can be included when one runs purely the LO flavour creation processes.

In general, Pythia is a leading-order event generator, but it includes some approximated NLO effects. For example the NLO flavour excitation production process for heavy quarks, where one heavy quark is kicked out of the proton, is partly treated as initial-state radiation in Pythia\footnote{In the EPOS3 case, the flavour excitation mechanism is modeled as a gluon splitting $ c\bar{c}$ pair in the ISR, with one of the heavy quarks undergoing a hard scattering in the Born process while the other stays in the projectile fragments.}. Due to these approximations for the NLO contributions, Pythia has problems to reproduce $Q\xoverline{Q}$ azimuthal correlations \cite{Acosta:2004nj,Khachatryan:2011wq,CMS:2016krr}. For our studies we use two approaches: a) the default LO+NLO Pythia 6 setting for the creation processes (MSEL=1) and b) the exclusively LO flavour creation processes (MSEL=4). One has to note here that it is impossible to normalise both approaches the same way because for the setup b) a $c\bar{c}$ pair is produced in each collision. A proper NLO treatment, as performed in Ref.~\cite{CMS:2016krr}, is left for future work. By applying a back-to-back selection of the heavy mesons, which is mostly sensitive to the leading-order production mechanism in the $t$-channel, we select the LO contributions in EPOS3+HQ. This allows for a meaningful comparison with the LO Pythia approach. The results from the CMS Collaboration give already a hint that the azimuthal correlations with a proper NLO treatment are similar to the LO Pythia approach for a back-to-back selection \cite{CMS:2016krr}.

\subsection{Calibration}
\label{subsect:Calibration}

Before starting to investigate the correlations between heavy quarks, we have to assure that the single particle distribution of the different approaches are comparable. This is a prerequisite for any correlation studies.

\begin{figure}[tb!]
    \begin{center}
    \includegraphics[width=0.9\columnwidth]{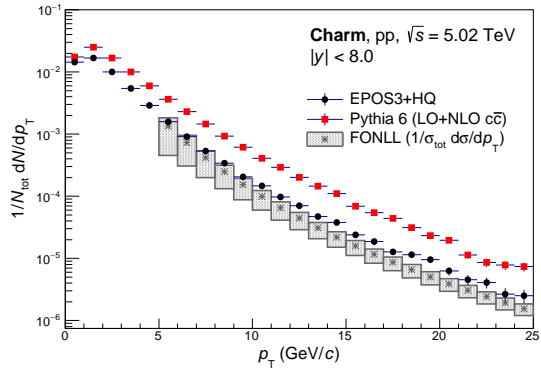}
    \end{center}
    \caption{Single particle spectra, ${\rm d}N/{\rm d}p_{\rm T}$, for charm quarks in pp collisions at $\sqrt{s} = 5.02$~TeV for EPOS3+HQ and the default Pythia 6 approach (LO+NLO $c\bar{c}$ production). The distributions are normalised to the number of events and compared to FONLL predictions \cite{Cacciari:1998it} (normalised using Ref.~\cite{Antchev:2013iaa}). See text for details.}
    \label{fig:calibration}
\end{figure}

In Figure~\ref{fig:calibration}, we present the single particle spectra ${\rm d}N/{\rm d}p_{\rm T}$ for charm quarks for EPOS3+HQ and the default LO+NLO Pythia 6 approach. The distributions are normalised to the total number of events. We see that EPOS3+HQ agrees well with the standard FONLL calculation \cite{Cacciari:1998it} whereas LO+NLO Pythia 6 overpredicts largely the FONLL predictions. This is due to NLO processes which are responsible for a large fraction of the $c\bar{c}$ pair production in LO+NLO Pythia \cite{Norrbin:2000zc}. Their importance varies strongly for different tunes. Therefore, other Pythia tunes may yield a better agreement with the FONLL total yield. For large $p_{\rm T}$ the slopes of the momentum distribution of both approaches are close to each other. This means that the different weighting of the different elementary $c\bar{c}$ pair creation processes in EPOS3 and LO+NLO Pythia yields about the same single particle spectrum at high $p_{\rm T}$. We analyse this further in section \ref{sec:azim}.

\section{Strategy}
\label{sec:Strategy}

\begin{figure}[tb!]
    \begin{center}
    \includegraphics[width=0.72\columnwidth]{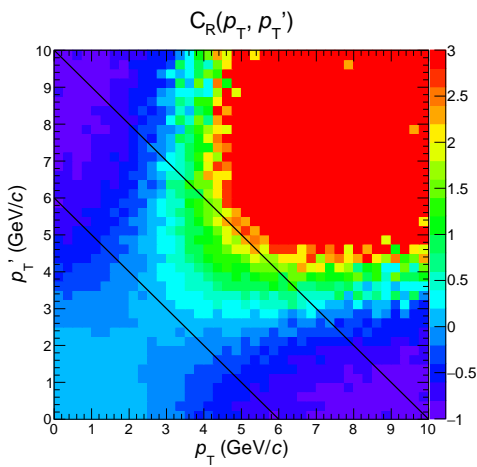}
    \end{center}
    \caption{The 2-dimensional correlation function $C_{R}( p_{\rm T}, p_{\rm T}{'} )$ for all back-to-back $c\bar{c}$ pairs in pp collisions at $\sqrt{s} = 5.02$~TeV generated with Pythia 6 (LO $c\bar{c}$ production). The two black lines indicate the selected $p_{\rm T}$ range for the final projection. Around the diagonal line at $\Delta p_{\rm T} = p_{\rm T} - p_{\rm T}{'} = 0$, a clear increase of the transverse momentum correlation is visible. Note that the maximum in the plot is set to 3 to have a reasonable colour scale.}
    \label{fig:CR_ccEPOS}
\end{figure}

For this study we are investigating the transverse momentum correlation of the $c\bar{c}$ pairs and their corresponding heavy-flavour mesons. We form all possible $\rm D\xoverline{\rm D}$ (or $c\bar{c}$) combinations in an event and employ an azimuthal selection: we concentrate on heavy meson pairs with a difference in the azimuthal angle of $\Delta \phi > 3\pi/4$ (back-to-back). As argued before, this is an important cut which allows to compare the results of EPOS3+HQ and Pythia 6. A $(p_{\rm T}, p_{\rm T}{'})$-contour plot displays the transverse momenta of all selected pair combinations in a sample, where $p_{\rm T}$ and $p_{\rm T}{'}$ are of the $\rm D$ and $\xoverline{\rm D}$ meson, respectively. From this contour plot, a 2-dimensional correlation function (independent of the grid size) is calculated, which is projected on the diagonal $\Delta p_{\rm T} = p_{\rm T} - p_{\rm T}{'}$ line for a specific $p_{\rm T} + p_{\rm T}{'}$ interval. This 1-dimensional projection is the final observable and will be compared for the different models. The single particle spectra ${\rm d}N/{\rm d}p_{\rm T}$ for these models are, as we will show, similar in the kinematical region where the primary $c$ and $\bar{c}$ quarks are emitted back-to-back. Comparison with experiment will therefore allow for constraining the models for heavy-quark energy loss and will reduce the uncertainties for the interpretation of pA and AA data.

In Figure~\ref{fig:CR_ccEPOS}, the 2-dimensional correlation function $C_{R}(p_{\rm T},p_{\rm T}{'})$ for $c\bar{c}$ pairs in pp collisions at $\sqrt{s}=5.02$~TeV, generated with Pythia 6, is displayed. The properties of these quarks compare with the so-called \textit{initial} stage in EPOS3+HQ. This initial stage includes the FSR process, but is taken before a possible interaction with a quark-gluon plasma, whereas the \textit{final} stage in EPOS3+HQ corresponds to the quarks just before hadronisation.

The calculation of the 2-dimensional correlation function $C_{R}(p_{\rm T},p_{\rm T}{'})$ is based on the fact that each two-dimensional function can be written as the product of two one-dimensional functions times a correlation function. First of all, we build the equivalent probability distribution:
\begin{equation}
    p( p_{\rm T}, p_{\rm T}{'} ) = \frac{ N( p_{\rm T}, p_{\rm T}{'} ) }{ \Delta p^2 \sum_{p_{\rm T}, p_{\rm T}{'}} N( p_{\rm T}, p_{\rm T}{'} ) } \, ,
    \label{eq:Equiprobdistr}
\end{equation}
where $N(p_{\rm T}, p_{\rm T}{'} )$ is the average number of counts in a $(p_{\rm T}, p_{\rm T}{'})$-cell and $\Delta p$ the grid size. The reduced probability distribution, which is basically a projection of the histogram on one of the two axes, is then given by:
\begin{equation}
    p_{R}(p_{\rm T}) = \Delta p \sum_{p_{\rm T}{'}} p( p_{\rm T}, p_{\rm T}{'} ) \, ,
    \label{eq:Reducedprobdistr}
\end{equation}
which satisfies $\Delta p \sum_{p_{\rm T}} p_{R}(p_{\rm T}) = 1$. Now the absolute correlation
\begin{equation}
  C(p_{\rm T}, p_{\rm T}{'}) = p(p_{\rm T}, p_{\rm T}{'}) - p_{R}(p_{\rm T})p_{R}(p_{\rm T}{'}) \, , 
\end{equation}
which is normalised to zero, and the relative correlation $C_{R}( p_{\rm T}, p_{\rm T}{'} )$ can be build:
\begin{equation}
  C_{R}(p_{\rm T}, p_{\rm T}{'}) = \frac{C(p_{\rm T}, p_{\rm T}{'})}{p_{R}(p_{\rm T}) p_{R}(p_{\rm T}{'})} \, .
  \label{eq:CR}
\end{equation}
Both, the absolute and the relative correlation, vanish in the absence of a correlation and are independent of the grid size (when $\Delta p$ is small enough and the statistics large enough). Note that using this method, the final correlation can be larger than one.

\section{Results}
\label{sec:Results}

For this paper, we study proton-proton collisions at $\sqrt{s} = 5.02$~TeV in both EPOS3+HQ and Pythia 6. This energy is chosen because there exist pp, p-Pb and Pb-Pb data, which offer the possibility to use this new observable to distinguish between different in-medium energy loss mechanisms (which we want to address in future work). It was checked that the message of this paper does not change when the energy is increased to $\sqrt{s} = 7$ or $13$~TeV.

The results at quark level will be shown first, because the picture is clearest there. For the correlations for the corresponding $\rm D$ mesons the additional effect of the fragmentation process comes into play. As the Peterson fragmentation function for charm quarks does peak at pretty large $x = p/p_{\rm max}\approx 0.8$ \cite{Peterson:1982ak}, information about the initial transverse momentum correlation will be mostly preserved when moving to the final-state $\rm D \xoverline{\rm D}$ mesons.

For the meson-meson transverse momentum correlations we study the combination of $\rm D^{+}$ or $\rm D^{0}$ with $\rm D^{-}$ and $\xoverline{\rm D^{0}}$ mesons. We are aware of the experimental difficulties to measure two fully reconstructed heavy-flavour mesons in one event, therefore we also studied (partly) heavy-flavour leptonic correlations and found similar results. These results are, however, not shown here because the leptonic decay introduces an additional distortion of the initial correlation.

\begin{figure}[t!]
    \begin{center}
    \includegraphics[width=0.75\columnwidth,trim={0 1.3cm 0 0},clip]{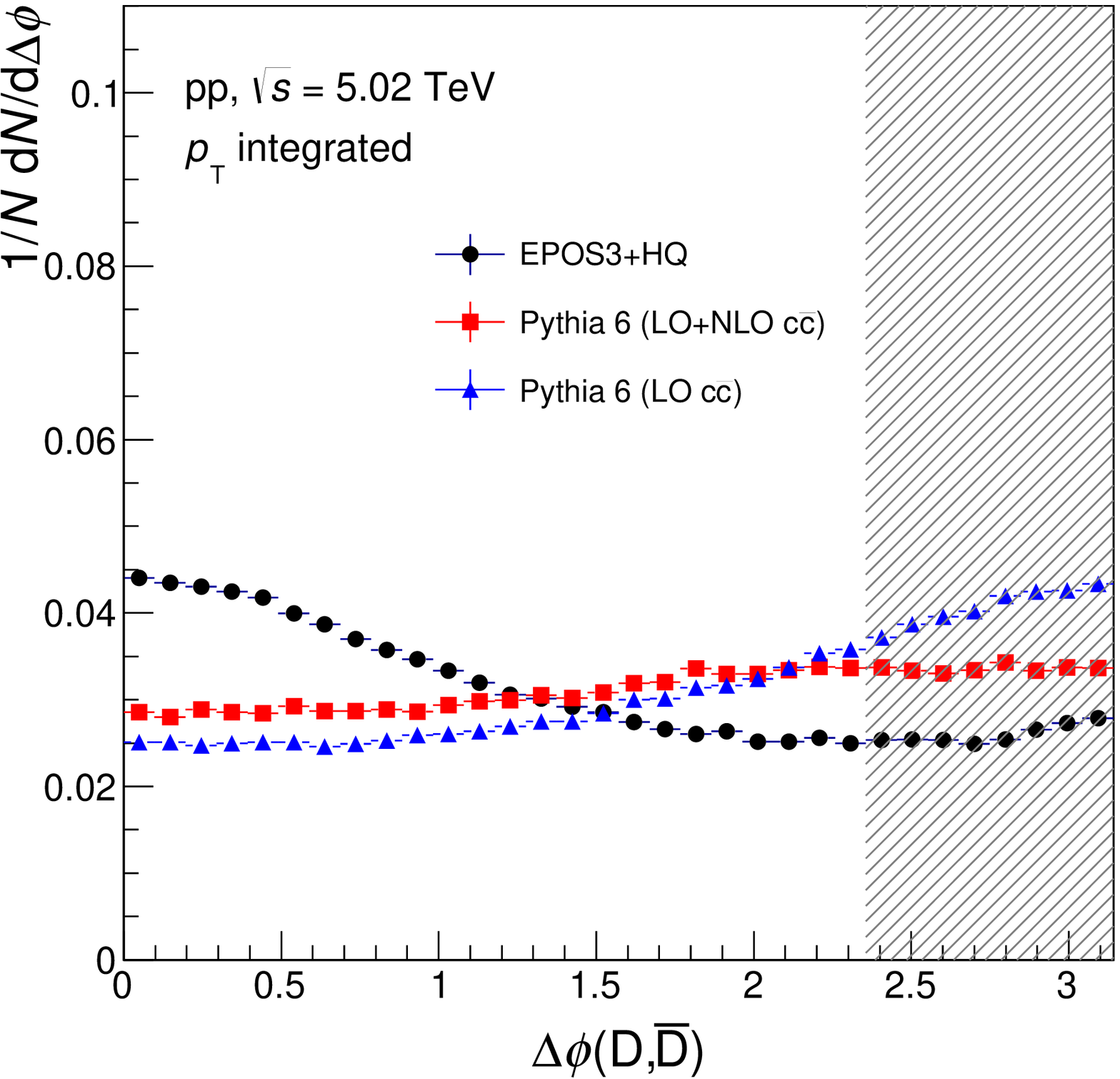}
    \includegraphics[width=0.75\columnwidth,trim={0 0 0 1.6cm},clip]{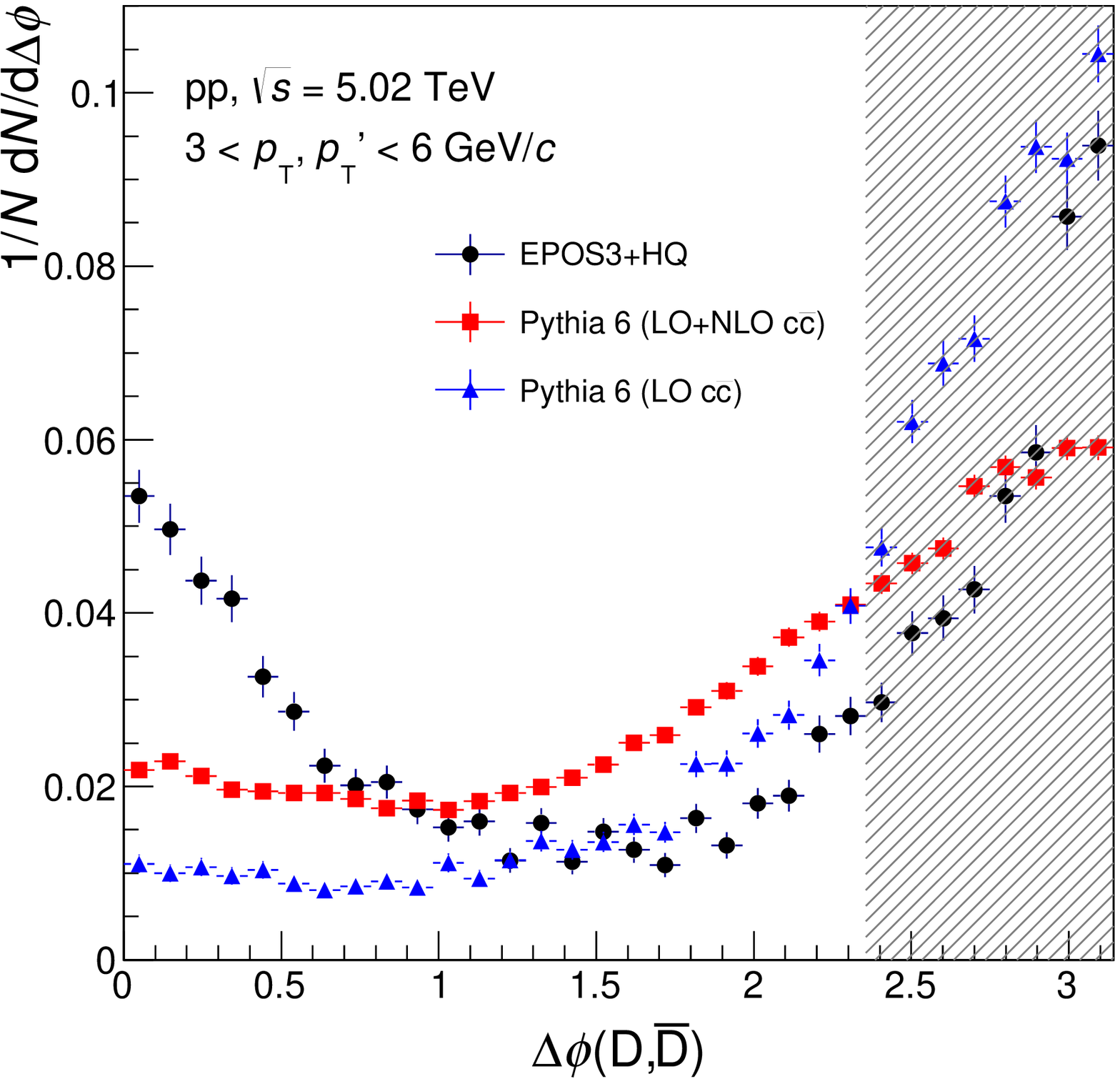}
    \end{center}
    \caption{The distribution of the difference in the azimuthal angle for all $\rm D\xoverline{\rm D}$ combinations in pp collisions at $\sqrt{s} = 5.02$~TeV for EPOS3+HQ and Pythia 6. The top figure shows the distributions irrespective of $p_{\rm T}$, where the selection $3 < p_{\rm T}, p_{\rm T}{'} < 6$~GeV/$c$ is applied in the bottom figure. The shaded band indicates the back-to-back selection cut. All distributions are normalised to one. See text for details.}
    \label{fig:dPhi_EPOS-PYTHIA}
\end{figure}

\subsection{Azimuthal angular correlations}
\label{sec:azim}

Only $\rm D\xoverline{\rm D}$ (or $c\bar{c}$) pairs that have a difference in azimuthal angle larger than $\Delta \phi > 3\pi/4$ are selected for this correlation study. The $\Delta \phi(\rm D,\xoverline{\rm D})$ distributions are depicted in Figure~\ref{fig:dPhi_EPOS-PYTHIA}, where one has to note that the default Pythia 6 distribution has problems reproducing measured $Q\xoverline{Q}$ azimuthal correlations \cite{Acosta:2004nj,Khachatryan:2011wq,CMS:2016krr}. The observed difference is expected, as we compare LO+NLO $c\bar{c}$ production processes for EPOS3+HQ with only LO $c\bar{c}$ in Pythia 6. For EPOS3+HQ we see clearly that the NLO processes, like gluon splitting, are dominant on the near side ($\Delta \phi(\rm D,\xoverline{\rm D}) = 0$). For Pythia, the leading-order peak at $\Delta \phi(\rm D,\xoverline{\rm D}) = \pi$ is clearly visible. If one selects a transverse momentum range (e.g. $3 < p_{\rm T}, p_{\rm T}{'} < 6$~GeV/$c$, see bottom Figure~\ref{fig:dPhi_EPOS-PYTHIA}), the back-to-back peak gets more dominant in both models. Besides that the $\Delta \phi > 3\pi/4$ selection is necessary for the model comparison, also the largest $p_{\rm T}$-$p_{\rm T}{'}$ correlation is expected for back-to-back pairs due to kinematic reasons.

The difference between the initial $\Delta \phi(c,\bar{c})$ and the final (before hadronisation) distributions is small in the EPOS3+HQ approach. This means that the influence of a possible QGP on the difference in the azimuthal angle is negligible. As we will see in the next subsection, the $p_{\rm T}$-$p_{\rm T}{'}$ correlation is affected more. The medium modifications of the azimuthal angular correlation distribution of heavy-flavour pairs in heavy-ion collisions were studied in Ref.~\cite{Song:2016rzw}.

\begin{figure}[tb!]
    \begin{center}
    \includegraphics[width=0.9\columnwidth]{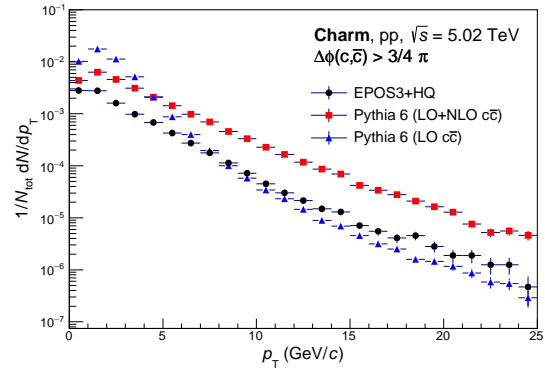}
    \end{center}
    \caption{Normalised single particle spectra, ${\rm d}N/{\rm d}p_{\rm T}$, for charm quarks with $\Delta \phi(c,\bar{c}) > 3\pi/4$ at $\sqrt{s} = 5.02$~TeV for EPOS3+HQ and both heavy-flavour production approaches for Pythia 6 (LO and LO+NLO $c\bar{c}$ production). See text for details.}
    \label{fig:calibration_b2b}
\end{figure}

The transverse momentum spectra for the back-to-back emitted charm quarks are shown in Figure~\ref{fig:calibration_b2b}. The EPOS3+HQ and LO+NLO Pythia curves are normalised to the number of events, whereas for the $c\bar{c}$ production in Pythia LO an additional normalisation factor has to be applied because there charm quarks are produced in each event. This factor is calculated in that way that the number of charm quarks with $p_{\rm T} > 15$~GeV/$c$ from LO flavour creation is similar for both Pythia 6 approaches. Above $p_{\rm T} = 6$~GeV a similar distribution (in the form as well as in the absolute value) is found for EPOS3+HQ and the LO $c\bar{c}$ Pythia 6 approach. Because we are interested in high momentum charm quarks the difference in the form of the distribution function at small $p_{\rm T}$ is not of concern for this comparison. In EPOS3+HQ this condition selects essentially LO processes. Thus, the necessary condition for a meaningful comparison of EPOS3+HQ and LO Pythia events is fulfilled. Regarding LO+NLO Pythia two observations are astonishing: a) the additional NLO processes, like flavour excitation and gluon splitting, do contribute substantially under the back-to back condition (about 87\% of the $c\bar{c}$ come from NLO events) and b) nevertheless the single particle heavy quark spectrum has for $p_{\rm T} > 10$~GeV almost the same form as that of LO Pythia. Hence, this new processes do not change the spectral shape and cannot be identified by measuring single particle spectra. Correlations, as also noted by Ref.~\cite{Field:2002da}, allow to discriminate between the different sources of heavy-flavour quarks.

\subsection{Transverse momentum imbalance}

\begin{figure}[tb!]
    \begin{center}
    \includegraphics[width=0.75\columnwidth]{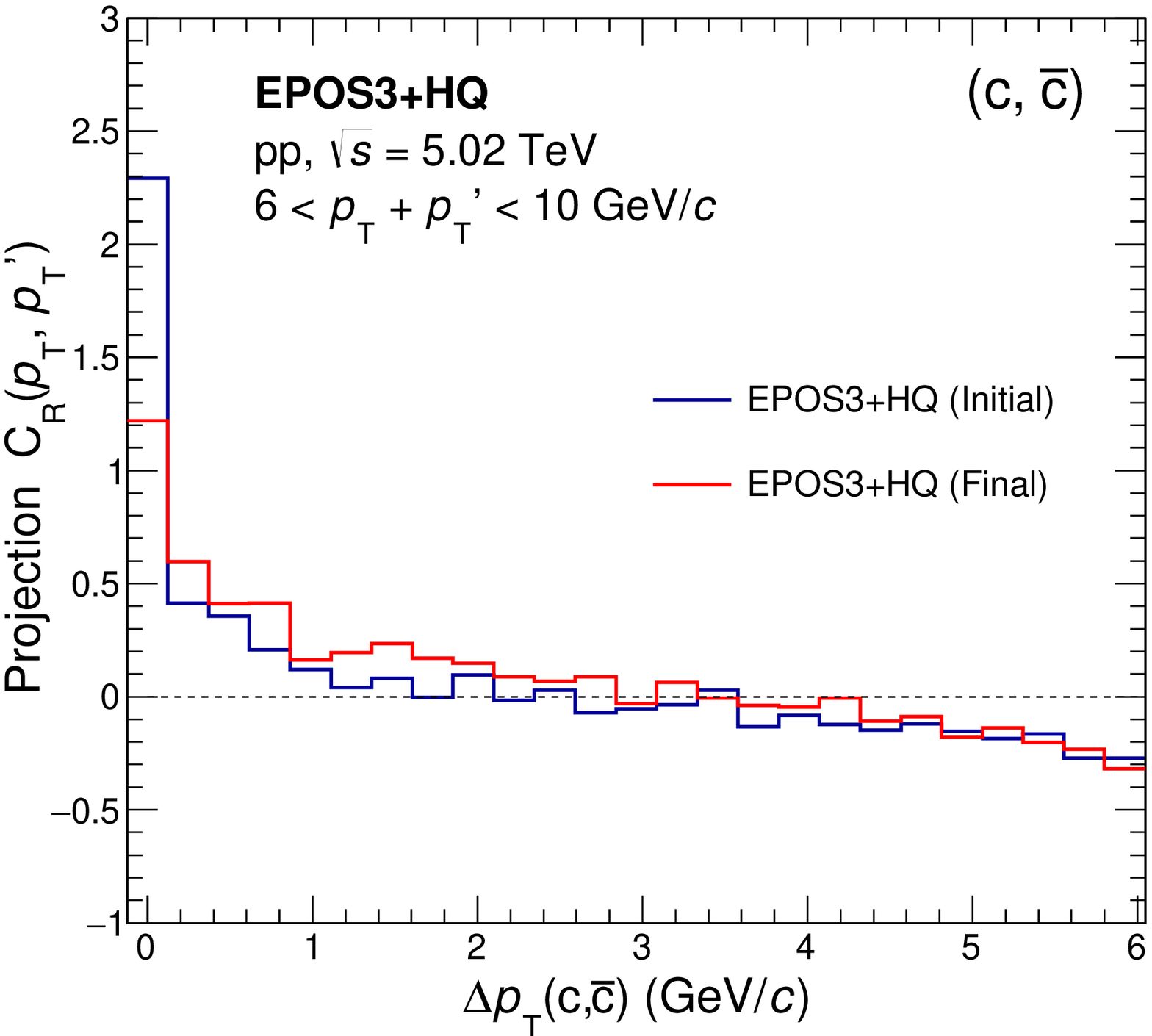}
    \includegraphics[width=0.75\columnwidth]{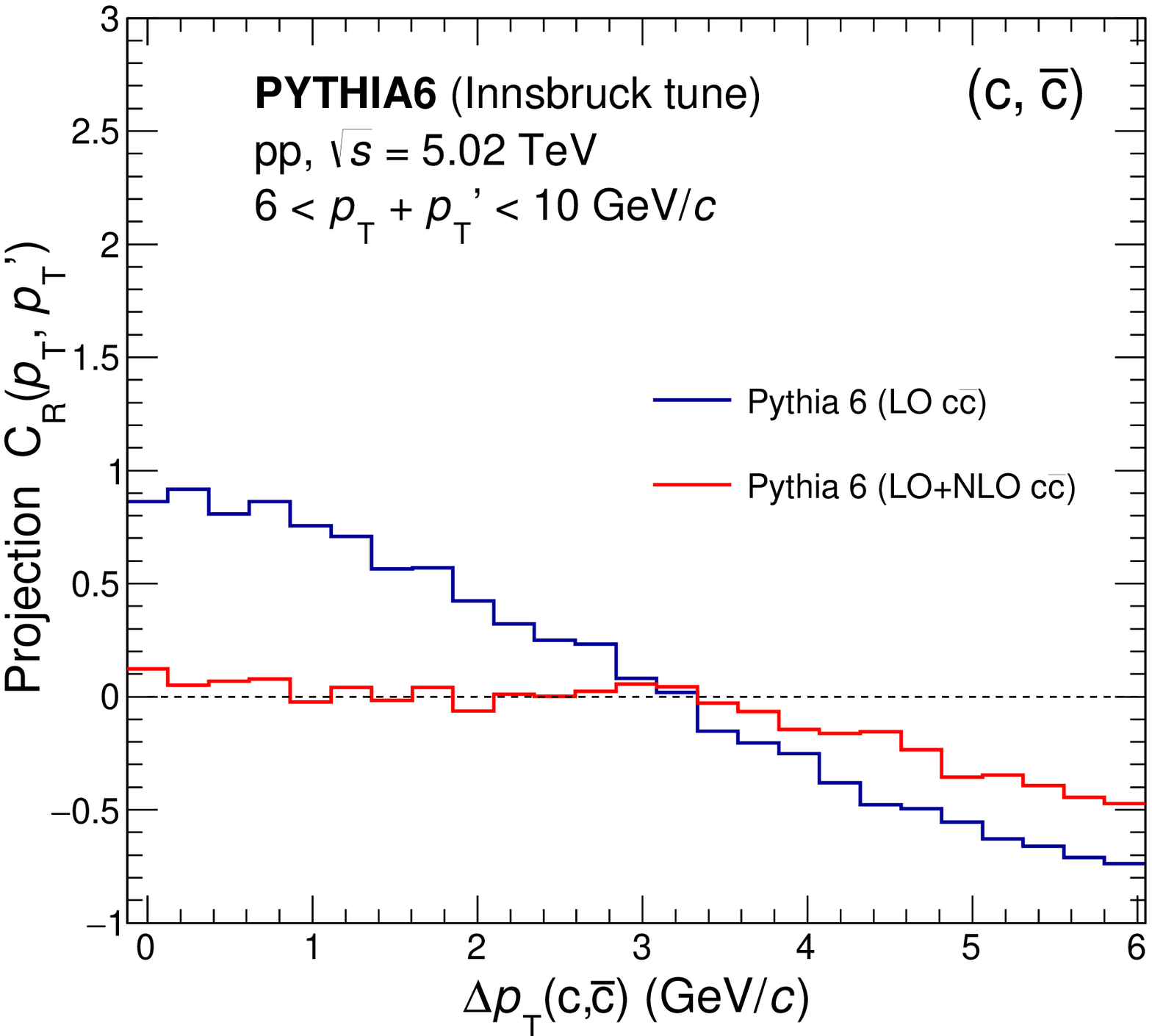}
    \end{center}
    \caption{The transverse momentum imbalance as a function of $\Delta p_{\rm T}$ for $c\bar{c}$ pairs with $\Delta \phi > 3\pi/4$ for EPOS3+HQ (top) and Pythia 6 (bottom) and $6 < p_{\rm T} + p_{\rm T}{'} < 10$~GeV/$c$. See text for details.}
    \label{fig:pTpTbar_ccbar_EPOS-PYTHIA}
\end{figure}

\begin{figure}[t!]
    \begin{center}
    \includegraphics[width=0.75\columnwidth]{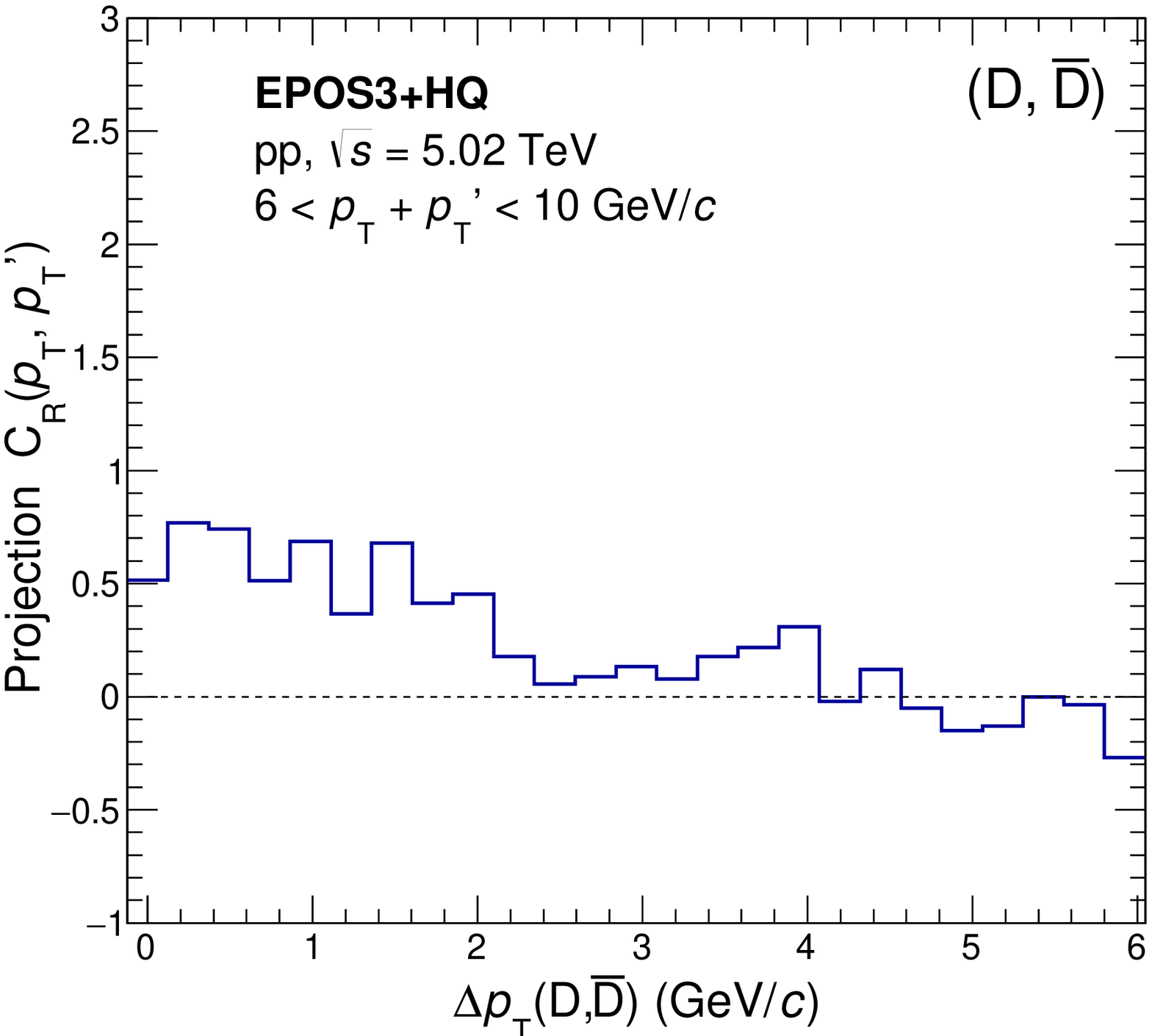}
    \includegraphics[width=0.75\columnwidth]{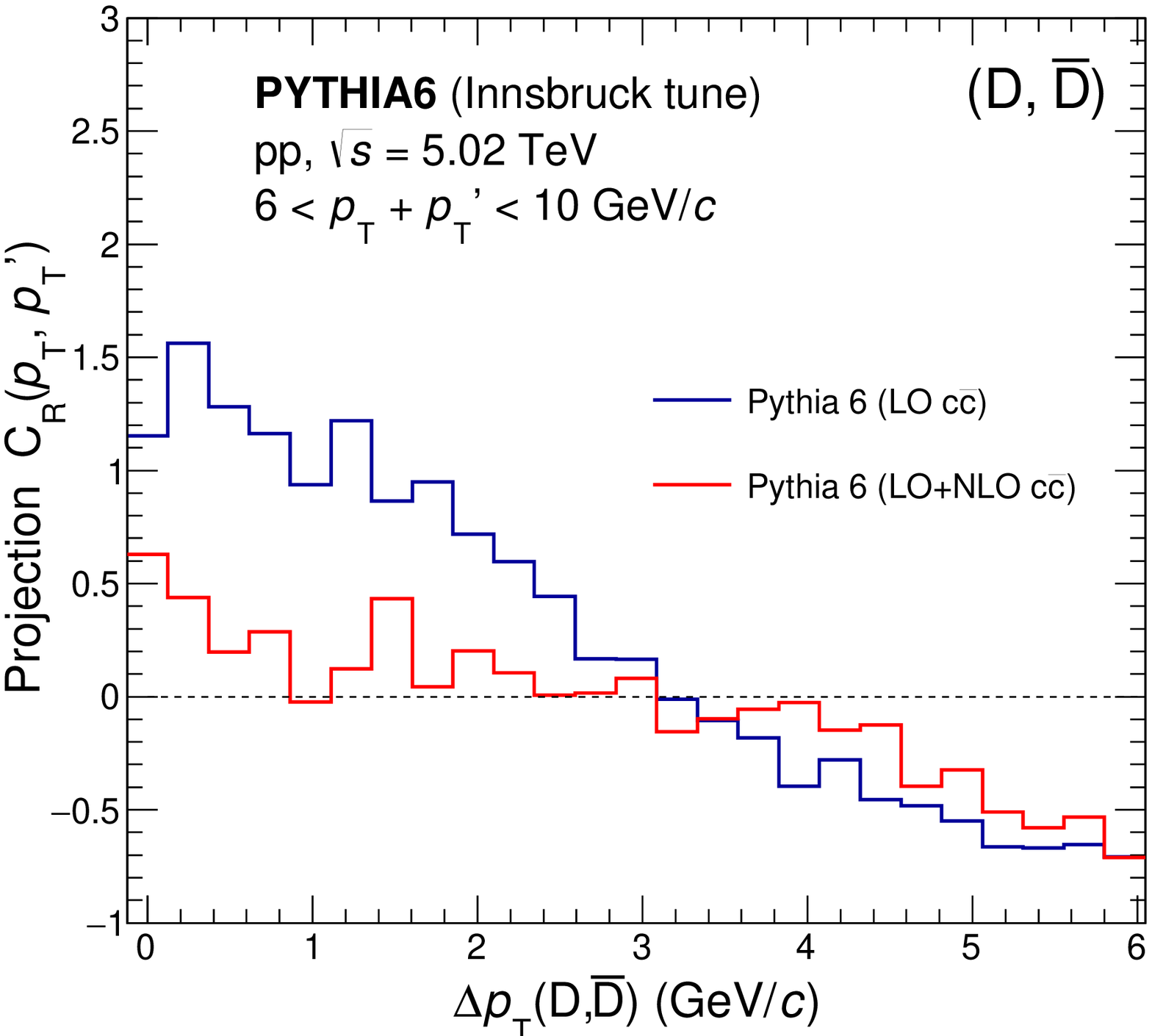}
    \end{center}
    \caption{The transverse momentum imbalance as a function of $\Delta p_{\rm T}$ for $\rm D\xoverline{\rm D}$ pairs with $\Delta \phi > 3\pi/4$ for EPOS3+HQ (top) and Pythia 6 (bottom) and $6 < p_{\rm T} + p_{\rm T}{'} < 10$~GeV/$c$. See text for details.}
    \label{fig:pTpTbar_DDbar_EPOS-PYTHIA}
\end{figure}

In Figure~\ref{fig:CR_ccEPOS}, we saw an example of the $C_{R}(p_{\rm T},p_{\rm T}{'})$ function. This 2-dimensional correlation function is projected on the diagonal $\Delta p_{\rm T}$ line for a specific $p_{\rm T} + p_{\rm T}{'}$ interval, defining our so-called transverse momentum imbalance observable. The transverse momentum imbalance for EPOS3+HQ and Pythia 6 are displayed in Figure~\ref{fig:pTpTbar_ccbar_EPOS-PYTHIA} (\ref{fig:pTpTbar_DDbar_EPOS-PYTHIA}) for $c\bar{c}$ ($\rm D\xoverline{\rm D}$) pairs. We choose for the diagonal band $6 < p_{\rm T} + p_{\rm T}{'} < 10$~GeV/$c$, which is relative large, in order to have sufficiently statistics. 

As shown in Figure~\ref{fig:pTpTbar_ccbar_EPOS-PYTHIA}, a clear transverse momentum correlation for $c\bar{c}$ pairs at $\Delta p_{\rm T} = 0$ is still visible after the final-state radiation process in EPOS3+HQ. Without in-medium interactions, we see $C_R \approx 2.3 $ for small $\Delta p_{\rm T}$ what means that such pairs are observed 3.3 times more often then expected if the momenta were uncorrelated. With in-medium interactions included, this maximal correlation value is smeared out. This is expected as collisions with medium particles will change the $p_{\rm T}$ of only one of the partners, which modifies the $p_{\rm T} = p_{\rm T}{'}$ symmetry. Also part of the heavy quarks will fall outside the total transverse momentum interval, as on average the heavy quarks loose more energy than they gain in interactions with a medium. So, the final EPOS3+HQ distribution in Figure~\ref{fig:pTpTbar_ccbar_EPOS-PYTHIA} corresponds to a slightly smaller sample than the initial one. Although the peak value at $\Delta p_{\rm T} = 0$ is decreased significantly, the overall shape of the correlation $C_{R}(p_{\rm T},p_{\rm T}{'})$ for $\Delta p_{\rm T} > 0$ stays the same.

$C_{R}(p_{\rm T},p_{\rm T}{'})$ as a function of the relative momentum $\Delta p_{\rm T}$ of the $c\bar{c}$ pairs for both Pythia 6 approaches is shown in the bottom panel of Figure~\ref{fig:pTpTbar_ccbar_EPOS-PYTHIA}. The correlation function for Pythia 6 does not show a clear enhancement at $p_{\rm T} = p_{\rm T}{'}$. For the LO $c\bar{c}$ sample, a slowly decreasing distribution, starting from $C_{R} = 0.9$, is found. For the Pythia sample using the default $c\bar{c}$ production mechanisms, the transverse momentum correlation is very small. Here, we should mention that the default LO+NLO settings in Pythia fails to reproduce the $\Delta\phi$ distributions because of the approximated NLO $c\bar{c}$ contributions \cite{Acosta:2004nj,Khachatryan:2011wq,CMS:2016krr}. Not only for small $p_{\rm T}$ values EPOS3+HQ and Pythia 6 differ, also at high $p_{\rm T}$ the behaviour is different. Whereas the LO $c\bar{c}$ distribution for Pythia 6 keeps decreasing, EPOS3+HQ stays almost flat and is close to zero.

Figure~\ref{fig:pTpTbar_DDbar_EPOS-PYTHIA} shows the transverse momentum imbalance for the final-state $\rm D$ mesons. The fragmentation process of the charm quarks broadens the peak at small values of $\Delta p_{\rm T}$. For both models, the overall shape at $\Delta p_{\rm T} > 0$ is similar for $c\bar{c}$ and $\rm D\xoverline{\rm D}$. A clear difference in both distributions, as already observed for $c\bar{c}$ pairs, is still visible.

\section{Conclusions}
\label{sec:Conclusions}

In this paper, we propose the imbalance of the transverse momentum for a given sum of the momenta of $c\bar{c}$ (${\rm D}\xoverline{\rm D}$) pairs as an observable which is sensitive to final-state radiation (FSR) of heavy quarks. Calculations using EPOS3+HQ and Pythia 6 were performed to study the effect of FSR. In EPOS3+HQ the formation of a plasma may have an additional (small) influence on the imbalance.

This new observable shows that the initial symmetric $p_{\rm T} = p_{\rm T}{'}$ correlation in EPOS3+HQ is not completely vanished by the final-state radiation processes, neither for the final $\rm D\xoverline{\rm D}$ nor for the $c\bar{c}$ before hadronisation. The condition that the heavy quarks are observed back-to-back selects quarks stemming from LO processes in EPOS3+HQ, which is observed by the agreement in slope of the single particle spectra for EPOS3+HQ and LO Pythia 6. Therefore, we can compare these two approaches directly and find that their momentum imbalance function differs considerably. EPOS3+HQ shows an enhancement for $\Delta p_{\rm T}=0$ but otherwise a rather flat distribution, whereas LO Pythia 6 does not show such an enhancement but a much stronger almost linear decrease as a function of $\Delta p_{\rm T}$. Consequently, the imbalance function allows to study experimentally which of these approaches gives the correct FSR. This observation is especially important to clarify the role of the dead cone effect for the FSR of heavy quarks.

The shape of this momentum imbalance of LO+NLO Pythia 6 differs also considerably from that of EPOS3+HQ as well as from that from LO Pythia 6. Both, EPOS3+HQ and LO+NLO Pythia 6, contain NLO contributions and so the momentum imbalance can shed light on the right weighting and the implementation of these production processes in the simulation programs.

Once fixed by comparing to measured pp data, this method can be extended to pA data where one can study how a cold medium effects the energy loss of heavy quarks. Finally, by comparing pp and AA collisions, we can study the influence of induced radiation \cite{Dokshitzer:2001zm,Aichelin:2013mra} as well as of that of elastic collisions of the heavy quarks with the medium. This we will address in a future paper. The presently available $R_{AA}$ and $v_{2}$ data have turned out not to be sufficient to define uniquely how the heavy quarks interact with the quark-gluon plasma. This new observable may therefore add to our understanding of the energy loss of heavy quarks in cold and hot nuclear matter.

{\bf Acknowledgments:} We acknowledge fruitful discussions with Matthew Nguyen and thank Taesoo Song and Elena Bratkovskaya to communicate to us the details of the Innsbruck tune. This work was supported by the region Pays de la Loire, the Netherlands Organisation for Scientific Research (project number: 680-47-232) and the Dutch Foundation for Fundamental Research (project numbers: 12PR3083).

\end{document}